\newif\ifDC \DCtrue
\ifDC
\documentclass[twocolumn,showpacs,preprintnumbers,amsmath,amssymb,prb]{revtex4}
\else
\documentclass[preprint,showpacs,preprintnumbers,amsmath,amssymb,prb]{revtex4}
\fi

\usepackage{graphicx}
\usepackage{dcolumn}
\usepackage{bm}

\newif\ifTEST \TESTfalse

\newcommand{\RH}{$R_{\rm H}$ }

\ifTEST
\newcommand{\citeChien}{[Ref. Chien]\cite{Chien91} }
\newcommand{\citeSegawa}{[Ref. Segawa]\cite{SegawaOverlap} }
\newcommand{\citeSegawaAnother}{[Ref. Segawa]\cite{SegawaOverlap} }
\newcommand{\citeSegawaZn}{[Ref. SegawaZn]\cite{SegawaZndoped} }
\newcommand{\citeIto}{[Ref. Ito]\cite{Ito} }
\newcommand{\citeAndoHall}{[Ref. AndoHall03]\cite{AndoHall03} }
\newcommand{\citeGagnon}{[Ref. Gagnon]\cite{Gagnon} }
\newcommand{\citeHusseyPRB}{[Ref. Hussey]\cite{HusseyPRB97} }
\newcommand{\citeAndoAnisotropy}{[Ref. AndoAnisotropy]\cite{AndoAnisotropy} }
\newcommand{\citeNoda}{[Ref. Noda]\cite{Noda} }
\newcommand{\citeMOSETC}{[Ref. MOSETC]\cite{AndoSNS,SegawaMOS02,YBCO_MR02} }
\else
\newcommand{\citeChien}{[Ref. 2]}
\newcommand{\citeSegawa}{[Ref. 20]}
\newcommand{\citeSegawaAnother}{20}
\newcommand{\citeSegawaZn}{[Ref. 18]}
\newcommand{\citeIto}{[Ref. 26]}
\newcommand{\citeAndoHall}{[Ref. 28]}
\newcommand{\citeGagnon}{[Ref. 13]}
\newcommand{\citeHusseyPRB}{[Ref. 29]}
\newcommand{\citeAndoAnisotropy}{[Ref. 15]}
\newcommand{\citeNoda}{[Ref. 36]}
\newcommand{\citeMOSETC}{[Refs. 33,37,38]}
\fi

\begin{document}

\title{Intrinsic Hall response of the CuO$_2$ planes
in a chain-plane-composite system of YBa$_2$Cu$_3$O$_y$}

\author{Kouji Segawa and Yoichi Ando}
\affiliation{Central Research Institute of Electric Power 
Industry, Komae, Tokyo 201-8511, Japan}

\date{November 14, 2003}
\begin{abstract}
The Hall coefficient is measured in YBa$_2$Cu$_3$O$_y$ untwinned
single crystals for a wide range of doping.
We show that the Hall conductivity and the Hall angle
of the {\it CuO$_2$ planes} in YBa$_2$Cu$_3$O$_y$
can be extracted from measurable transport properties
regardless of the conduction of the Cu-O chains
nor the in-plane anisotropy of the CuO$_2$ planes.
The present analysis allows us to discuss
the genuine Hall effect in the CuO$_2$ planes
alone in YBa$_2$Cu$_3$O$_y$
without any complications due to the Cu-O chains.
\end{abstract}

\pacs{74.25.Fy, 74.72.Bk}

\maketitle

\newlength{\figurewidthwide}
\newlength{\figurewidthnarrow}
\ifDC
\setlength{\figurewidthwide}{8.5cm}
\setlength{\figurewidthnarrow}{6cm}
\else
\setlength{\figurewidthwide}{13cm}
\setlength{\figurewidthnarrow}{9cm}
\fi

\section{Introduction}
The origin of the peculiar normal-state properties
of the high-$T_{\rm c}$ superconductors is not fully understood.
The strongly temperature-dependent Hall coefficient ($R_{\rm H}$)
is one of the best-known peculiar features in the normal state
of high-$T_{\rm c}$ cuprates \cite{OngReview}.
While $R_{\rm H}$ shows a complicated temperature dependence,
the cotangent of the Hall angle ($\cot \Theta_{\rm H}\equiv E_x/E_y=\rho_{xx}/\rho_{yx}$)
was reported to show a simple $T^2$-dependence in YBa$_2$Cu$_3$O$_y$ (YBCO)
\citeChien;
it is widely believed that
the peculiar behavior of \RH comes from
an anomalous coexistence of the $T^2$-law in $\cot \Theta_{\rm H}$
and the pronounced $T$-linear resistivity.
This fact suggests that two kinds of scattering times
dominate charge transport in high-$T_{\rm c}$ cuprates
\cite{Anderson91,Carrington92,Alexandrov94,Kotliar96,Coleman96,Stojkovic97,Ioffe98,Zheleznyak99,Kontani99,Varma01}.

However, the YBCO system has the Cu-O chains which are likely
to show one-dimensional (1D) electronic conduction \cite{Gagnon},
which is believed to cause the in-plane resistivity anisotropy
in the highly-doped region \cite{Gagnon,Takenaka,AndoAnisotropy}.
On the other hand, in the underdoped region, the 1D conduction of the Cu-O chains
is expected to diminish quickly with reducing oxygen content,
because 
oxygens are removed from the Cu-O chains, creating vacancies in the chains, and
1D systems are known to be very sensitive to disorder \cite{Hoffmann}.
However, we have demonstrated that the in-plane anisotropy does not disappear
in the underdoped region, which indicates that the CuO$_2$ plane itself is anisotropic \cite{AndoAnisotropy}.
In such a material with in-plane anisotropy as well as the chains,
resistivity and $\cot \Theta_{\rm H}$ measured in twinned crystals are mixtures of the intrinsic transport properties.
The original report of the $T^2$-law in $\cot \Theta_{\rm H}$ \citeChien{}
was based on the result in twinned single crystals,
and therefore the conductive Cu-O chains may have affected the temperature dependences
of $\cot \Theta_{\rm H}$ and resistivity.
Detwinning the crystals allows us to observe the in-plane anisotropy in resistivity;
however, even the results in untwinned single crystals \cite{JPRice} are not
free from contribution of the Cu-O chains,
because the Cu-O chains can indirectly modify the Hall coefficient, as described later.
Therefore, measurement on untwinned crystals
and estimating the contribution from the Cu-O chains
are necessary to obtain intrinsic transport properties
in YBCO in the whole doping region.

In this paper, we report the way to deduce the genuine Hall response
of the {\it CuO$_2$ planes}
in YBa$_2$Cu$_3$O$_y$
untwinned single crystals.
We show that the Hall conductivity ($\sigma_{xy}$)
and the cotangent of the Hall angle ($\cot \Theta_{\rm H}$)
of the {\it CuO$_2$ planes} can be obtained from measurable
transport properties such as the $a(b)$-axis in-plane resistivity $\rho_{a(b)}$
and the Hall coefficient $R_{\rm H}$.
The result shows that the electronic state of the CuO$_2$ planes
is responsible for the observed `60-K anomalies' in YBCO, such as
a decrease in the Hall coefficient just above $T_{\rm c}$
and the enhancement of the Hall conductivity and the Hall mobility
in samples with $y\sim 6.65$.

\section{Experiments}

The YBCO single crystals are grown in Y$_2$O$_3$ crucibles
by a conventional flux method \cite{SegawaZndoped}.
%
A wide range of oxygen contents are achieved
by annealing under various conditions
as shown in Table \ref{Annealing_conditions}
and quenching from the set temperatures
of the annealing.
Slightly overdoped ($y\simeq 7.00$)
and optimally doped ($y\simeq 6.95$) crystals
are obtained by annealing in oxygen atmosphere for a long time.
Samples with $y=6.70$--6.85
are obtained by annealing in air
for 12--48 hours at temperatures which increase
with decreasing oxygen contents.
For more underdoped samples,
annealing in air
is not applicable because
it is difficult to
quench samples quickly enough for preventing
absorption of oxygen onto the surface of the crystals
at very high temperatures.
Instead, in order to obtain oxygen contents
of $y\le 6.65$ we seal crystals in a quartz tube
together with polycrystalline powders with controlled oxygen contents.
The amount of the buffer powder should be as large as possible
for obtaining homogeneous oxygen distribution;
however,
if the amount of powder is too large (e.g., more than 100 mg in $\sim$7
c.c. sealed quartz tube)
we observe superconducting transition of minor phases
at around 60 K.
Therefore, we use $\sim$50 mg of powder for each annealing,
keeping the amount of single crystals less than 1\% of powder.
Only samples with $y=6.30$ are annealed by using
special furnace in which we can control the oxygen partial pressure.
We keep samples at a room temperature for at least a week after any heat treatments,
because the room-temperature annealing effect \cite{Lavrov}
is observed in the time scale of a few days.
%
All crystals are detwinned
at $\le$ 210 $^\circ$C in flowing nitrogen {\it after} the annealing,
because annealing at high temperatures above $\sim$600 $^\circ$C
destroys an untwinned state.
We confirmed that the experimental results
are not affected by the order of annealing and detwinning
for samples with $y= 6.80$ and 6.85 \citeSegawa, which can be detwinned
either before or after the annealing.
We decrease the detwinning temperature with decreasing oxygen contents;
for example, the detwinning temperature for $y=7.00$ samples is $\sim$210 $^\circ$C,
and that for $y=6.30$ samples is $\sim$120 $^\circ$C under
an uniaxial pressure of $\sim$0.1 GPa.
The oxygen content $y$ is determined by iodometric titration \cite{Kishio}
on powders which are annealed with the crystals,
because the volume of the crystals themselves is too small for the titration.
The error in $y$ is $\pm 0.02$ \citeSegawaZn.

\newcommand{\TableOne}{
\begin{table}
\begin{center}
\begin{tabular}{cccccc}
\hline
$y$ & $T_{\rm anneal}$ & keeping & atmosphere & $T_{\rm quench}$ ($^{\circ}$C) & $T_{\rm c}$ \\
& ($^{\circ}$C) & time & & of powder & (K) \\ \hline
7.00 & 400 & $\ge 10$ days & $\sim 2$ atm O$_2$ & -- & 91 \\
6.95 & 485 & $\ge 7$ days & 1 atm O$_2$ & -- & 93 \\
6.85 & 535 & 48 hours & air & -- & 83 \\
6.80 & 565 & 24 hours & air & -- & 69 \\
6.75 & 600 & 12 hours & air & -- & 62 \\
6.70 & 625 & 12 hours & air & -- & 60 \\
6.65 & 550 & 24 hours & sealed & 620 & 58 \\
6.60 & 600 & 12 hours & sealed & 640 & 57 \\
6.55 & 600 & 12 hours & sealed & 660 & 55 \\
6.50 & 600 & 12 hours & sealed & 675 & 35 \\
6.45 & 600 & 12 hours & sealed & 700 & 20 \\
6.35 & 700 & 6 hours & sealed & 800 & -- \\
6.30 & 550 & 36 hours & $3.5\times 10^{-4}$ atm O$_2$ & -- & -- \\ \hline
\label{Table1}
\end{tabular}
\end{center}
\caption{Annealing conditions for our YBCO crystals.
$T_{\rm quench}$ is the temperature
at which polycrystalline YBCO powders are quenched in air, and
the polycrystalline powders are sealed in a quartz tube together with
YBCO single crystals for tuning the oxygen contents.}
\label{Annealing_conditions}
\end{table}}
\ifDC
\TableOne
\fi

The Hall coefficient data are taken by sweeping the magnetic field
(along the $c$-axis) to both plus and minus polarities
up to 10--14 T
at fixed temperatures.
For each composition,
the $R_{\rm H}$ data are presented only for those temperatures at which the 
Hall voltage is perfectly proportional to the magnetic field \cite{AbeZnYBCO}.
Figure \ref{B_dependence_Figure} shows the field dependences of the Hall resistivity
for an optimally doped sample at various temperatures.
The origin for each temperature is shifted along $y$-axis
by 0.5 $\mu \Omega$ cm for 93--100 K and 0.25 $\mu \Omega$ cm for 100--300 K
for clarity.
We cannot extract $R_{\rm H}$ from the data for 93 and 95 K,
because the field dependences
deviate from the $H$-linear behavior by superconducting fluctuation.

\newcommand{\FigureOne}{
\begin{figure}
\includegraphics[width=\figurewidthnarrow]{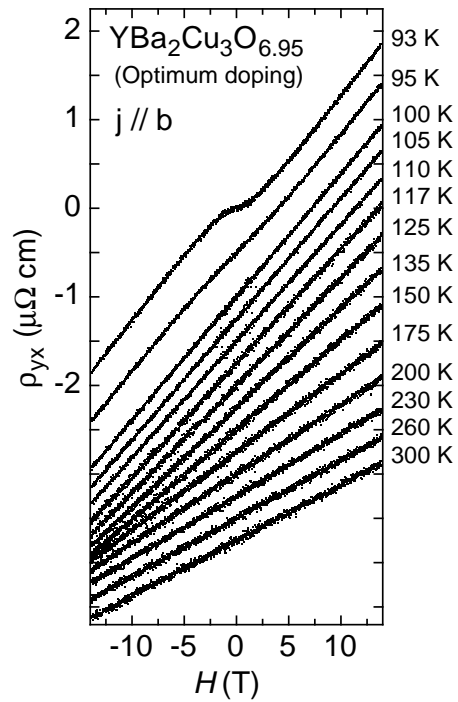}
\caption{Magnetic-field dependence of $\rho_{yx}$ for a YBCO crystal
with $y$=6.95 (optimum doping) at various temperatures.}
\label{B_dependence_Figure}
\end{figure}}
\ifDC
\FigureOne
\fi

\section{Experimental Results}
\subsection{In-plane anisotropy of the Hall coefficient}
According to the Onsager's reciprocal relation \cite{Onsager}
[$\rho_{yx}(H)=\rho_{yx}(-H)$],
when the magnetic field is applied along the $c$-axis,
$R_{\rm H}$ measured
in a sample with the current along the $a$-axis ($\rho_a$-sample)
is the same as that
with the current along the $b$-axis ($\rho_b$-sample)
unless time-reversal symmetry is broken.
However, YBCO is a complicated composite system containing
two-dimensional (2D) CuO$_2$ planes and one-dimensional (1D)
Cu-O chains and thus it is not self-evident
whether the Onsager relation is satisfied in YBCO.
Harris {\it et al.} mentioned that the Onsager relation
was accurately satisfied \cite{Harris95},
but no actual data were shown in their paper.
Figure \ref{Onsager_Figure} shows the temperature dependences
of $R_{\rm H}^{a}$ and $R_{\rm H}^{b}$ in an optimally doped sample,
which were measured with the current along the $a$-axis and the $b$-axis;
it should be noted that these data were measured for the same sample
with the same electrodes
by re-detwinning the crystal without changing the oxygen content.
Therefore, we have eliminated the uncertainty
in the estimation of the sizes of the crystal and electrodes from comparison.
The two curves shown in Fig. \ref{Onsager_Figure} do not coincide but are very close to each other.
$R_{\rm H}^{b}$ is slightly (at most $\sim$ 4 \%) smaller than $R_{\rm H}^{a}$,
but this difference is most likely due to an extrinsic effect of the
Hall measurement on a finite-sized sample
combined with the resistivity anisotropy \cite{Note}.
Therefore, we can say that the Onsager relation indeed holds
within an error of $\sim$ 5\%.

\newcommand{\FigureTwo}{
\begin{figure}
\includegraphics[width=\figurewidthnarrow]{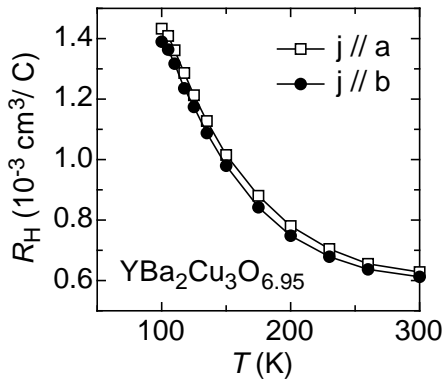}
\caption{
$R_{\rm H}(T)$ for two measurements
on the same crystal
at $y$=6.95 (optimum doping),
with $j\parallel a$(open squares) and $j\parallel b$ (solid circles).}
\label{Onsager_Figure}
\end{figure}}
\ifDC
\FigureTwo
\fi

\subsection{Measured Hall coefficient}

Figure \ref{measured_RH_Figure} shows the temperature dependences of the measured
Hall coefficient
($R_{\rm H}^{\rm meas}$)
for samples with various oxygen contents in a semi-log plot.
In general, the magnitude of $R_{\rm H}^{\rm meas}$ increases
with decreasing oxygen content.
At high temperatures for most samples $R_{\rm H}^{\rm meas}$
increases with decreasing temperature, while at low temperatures
various behaviors are observed.
For slightly underdoped samples with $y$=6.60--6.85,
$R_{\rm H}^{\rm meas}$ shows a decrease at low temperatures
below $\sim$ 120 K unlike highly doped samples with $y$=6.95--7.00.
The observed decrease appears to be the strongest in samples with $y$=6.75--6.80.
These results are roughly consistent
with the previously published data for twinned crystals with $y\ge 6.45$
\citeIto.
For less doped samples the low-temperature decrease
becomes weaker and eventually almost constant $R_{\rm H}^{\rm meas}$
is observed at low temperatures for samples with $y$=6.45--6.55.
In the lightly-doped nonsuperconducting samples with $y=6.30$ and 6.35,
$R_{\rm H}^{\rm meas}$ increases with decreasing temperature
at low temperatures, which is due to the charge localization observed in
the in-plane resistivity \cite{AndoAnisotropy}.

\newcommand{\FigureThree}{\begin{figure}
\includegraphics[width=\figurewidthnarrow]{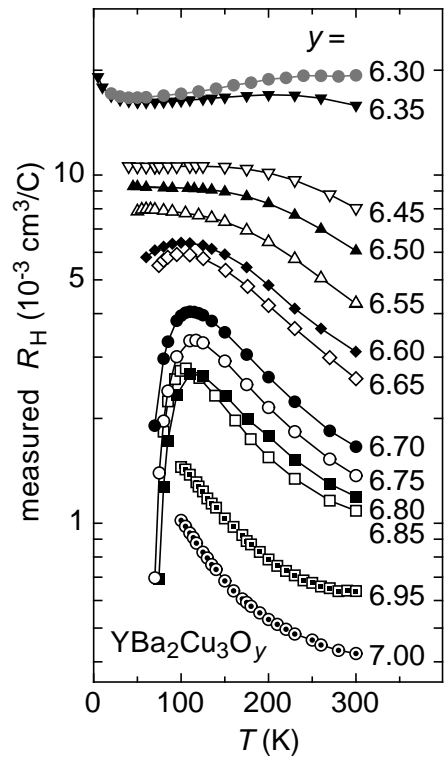}
\vspace{0.2cm}
\caption{$T$ dependences of $R_{\rm H}^{\rm meas}$ in untwinned YBCO crystals%
where the magnetic field is along the $c$-axis.}
\label{measured_RH_Figure}
\end{figure}}
\ifDC
\FigureThree
\fi

In the lightly doped region, nearly constant $R_{\rm H}^{\rm meas}$
may allow us to discuss carrier concentrations extracted from the Hall
coefficients.
Figure \ref{n_Figure} shows the temperature dependence of the apparent carrier concentration
which is calculated by $V/NeR_{\rm H}^{\rm meas}$, where
$V$ is the volume of a unit cell and $N$ is the number of
Cu atoms per unit cell \cite{AndoCarrierConc}.
The data show plateaus at low temperatures, where values of
$V/NeR_{\rm H}^{\rm meas}$ tell us
the actual carrier concentrations;
we can estimate the carrier concentration of the samples
with $y$=6.30(6.35), 6.45, 6.50 and 6.55 to be $\sim 3\%$, $\sim 5\%$, $\sim 6\%$
and $\sim 7\%$,  respectively.
Validity of this method in estimating the carrier concentration
is confirmed in a recent experiment in LSCO \citeAndoHall.
The carrier concentration at the superconductor-insulator boundary in YBCO
appears to be a reasonable value of  $\sim 4.5$\% and thus
the Hall coefficient in this region 
is apparently not affected by the conduction of the Cu-O chains.
However, for more doped samples
\RH shows the complicated temperature dependence,
which is possibly modified by
a contribution of the Cu-O chains
if the Cu-O chains are conductive.
Next we try to extract the Hall coefficient in the CuO$_2$ planes
for samples with possibly conductive Cu-O chains.

\newcommand{\FigureFour}{
\begin{figure}
\includegraphics[width=\figurewidthnarrow]{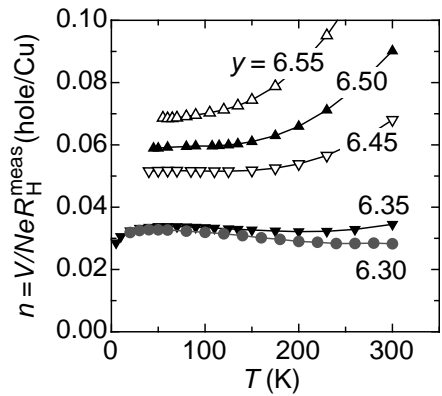}
\vspace{0.2cm}
\caption{$T$ dependences of $V/NeR_{\rm H}^{\rm meas}$ in untwinned YBCO crystals
with $y=6.30$-6.55.
}
\label{n_Figure}
\end{figure}}
\ifDC
\FigureFour
\fi

\section{Analysis and Discussion}
In this section, we extract the physical properties of the CuO$_2$ planes
by evaluating the effect of 1D conduction of the Cu-O chains.
Our analysis is based on the
parallel resistor model (PRM) along the $b$-axis \cite{Gagnon},
by which a $T^2$-law
in the resistivity of the Cu-O chains
was successfully extracted
in optimally doped YBCO \citeGagnon{}
and YBa$_2$Cu$_4$O$_8$ (Y124) \citeHusseyPRB.
Excellent $T^2$-law is also observed
in the resistivity of the Cu-O chains
($\rho^{\rm ch}$)
in our slightly overdoped samples with $y$=7.00 as shown in Fig. \ref{Fig_y700}(a).
In addition, the value of the residual resistivity for $\rho^{\rm ch}$
(35.7 $\mu \Omega {\rm cm}$)
is roughly a third of
the previously reported value \cite{Gagnon}, which evinces the
cleanliness of our crystals.
In the PRM
we assume that
(1) the Cu-O chains show electronic conduction along their direction,
(2) inter-chain conduction is negligible, and
(3) conduction bands of the Cu-O chain layer and the CuO$_2$ planes do not mix.
Since the Cu-O chains run along the $b$-axis,
the total conductivity along the $b$-axis is
\[
\sigma_b^{\rm tot}=\sigma^{\rm pl}+\sigma^{\rm ch},
\]
where $\sigma^{\rm pl}$ and $\sigma^{\rm ch}$
are the conductivity of the CuO$_2$ planes and the Cu-O chains, respectively.
On the other hand,
the total conductivity along the $a$-axis is
\[
\sigma_a^{\rm tot}=\sigma^{\rm pl},
\]
since the conduction perpendicular to the Cu-O chains is neglected.
The in-plane anisotropy of the electrical conduction of the CuO$_2$ planes
is neglected for the moment
(we will discuss the case of anisotropic CuO$_2$ planes later).

\subsection{Hall coefficient}

First we extract the Hall coefficient of the CuO$_2$ planes
in the $\rho_b$-sample because the process is simpler than that in the $\rho_a$-sample.
In the $\rho_b$-sample the current flows separately through the CuO$_2$ planes and the Cu-O chains
so that the current flowing through the CuO$_2$ planes ($j_x^{\rm pl}$) is given by
\[
j_x^{\rm pl}  = j_x^{\rm tot} \left( \frac{\sigma^{\rm pl}}{\sigma^{\rm pl}+\sigma^{\rm ch}}\right)
= j_x^{\rm tot} \left( \frac{\sigma_a^{\rm tot}}{\sigma_b^{\rm tot}}\right)
= j_x^{\rm tot} \left( \frac{\rho_b}{\rho_a}\right),
\]
\noindent
where $j_x^{\rm tot}$ is the total current.
In the PRM it is expected that the Hall voltage is produced only by this current $j_x^{\rm pl}$
and the Cu-O chains do not affect the Hall electric field.
(Although some finite Hall coefficient is observed in quasi-1D systems
\cite{Horii02,Mihaly00},
as we discuss in Section \ref{quasi1D},
contribution from the Cu-O chains to \RH
can be neglected in the case of YBCO.
Therefore, the Hall coefficient of the CuO$_2$ planes
which are assumed to be isotropic, $R_{\rm H}^{\rm{pl(iso)}}$,
is calculated as
\begin{equation}
R_{\rm H}^{\rm pl(iso)}  = R_{\rm H}^{\rm meas} \left(\frac{j_x^{\rm tot}}{j_x^{\rm pl}}\right) =
R_{\rm H}^{\rm meas} \left(\frac{\rho_a}{\rho_b}\right).
\label{HallDef_isotropic}
\end{equation}

The same result is obtained also in the $\rho_a$-sample.
The longitudinal current $j_x^{\rm tot}$ flows only through the CuO$_2$ planes
in the $\rho_a$-sample, and thus $j_x^{\rm pl}=j_x^{\rm tot}$ and $j_x^{\rm ch}$=0
are satisfied along the $a$-axis.
Naively, the measured Hall coefficient itself may seem to signify the Hall coefficient of the CuO$_2$ planes;
however, the observed Hall voltage is reduced by the short-circuiting effect of
the Cu-O chains in the $\rho_a$-sample,
because the transverse electric field $E_y$
takes effect not only in the CuO$_2$ planes but also in the Cu-O chains.
Since $E_{y}$ is parallel to the Cu-O chains,
a current $j_y^{\rm ch}$
inevitably flows through the Cu-O chains.
Therefore, in the Cu-O chains
\[ E_y= \rho^{\rm ch}j_y^{\rm ch} \]
is satisfied.
On the other hand,
a counter current of the same magnitude of $j_y^{\rm ch}$
must flow through the CuO$_2$ planes
because of the condition $j_y^{\rm tot}$=0,
where $j_y^{\rm tot}$ is the transverse current of the total system along the $y$-axis.
Therefore, $E_y$ in the CuO$_2$ planes is calculated as
\[ E_y=E_{\rm H} - \rho^{\rm pl}j_y^{\rm ch}, \]
where $E_{\rm H}$ is the Hall electric field which
is generated to compensate the Lorentz force
and $E_{\rm H}$ is given by
\[ E_{\rm H}=\rho_{yx}^{\rm pl(iso)}j_x^{\rm pl},\]
where $\rho_{yx}^{\rm pl(iso)}$ is the Hall resistivity of the isotropic CuO$_2$ planes.
Since $E_y$ is identical in the CuO$_2$ planes and in the Cu-O chains,
\[ E_{\rm H} - \rho^{\rm pl}j_y^{\rm ch} = \rho^{\rm ch}j_y^{\rm ch}\]
is satisfied.
We can express $j_{y}^{\rm ch}$ by using measurable properties as
\[ j_{y}^{\rm ch}=\frac{E_{y}}{\rho^{\rm ch}}
=E_{y}\left(\frac{1}{\rho_b}-\frac{1}{\rho_a}\right). \]
Therefore, we obtain
\[ E_{\rm H}=E_{y}+\rho_a j_{y}^{\rm ch}
= E_{y}\left(\frac{\rho_a}{\rho_b}\right), \]
and by dividing both sides of this equation by $j_x^{\rm tot}(=j_x^{\rm pl})$
Eq. (\ref{HallDef_isotropic}) is obtained (using $\rho_{yx}^{\rm pl(iso)} = E_{\rm H}/j_x^{\rm pl}$
and $\rho_{yx}^{\rm meas}=E_{y}/j_x^{\rm tot}$).

In the above analysis, the anisotropy of
the resistivity in the CuO$_2$ planes is ignored.
In other words, the global in-plane anisotropy
is assumed to be caused only by the Cu-O chains.
In recent measurements, however, we observed
considerable anisotropy in the in-plane resistivity
in oxygen deficient YBCO \citeAndoAnisotropy.
It is unlikely that oxygen deficient Cu-O chains
show good 1D conduction,
and thus the CuO$_2$ planes themselves must be anisotropic
in the underdoped YBCO.
If we re-calculate Eq. (\ref{HallDef_isotropic}) under the assumption
of the anisotropic CuO$_2$ planes, we obtain
\begin{equation}
R_{\rm H}^{\rm pl}=
R_{\rm H}^{\rm meas} \left(\frac{\rho_a}{\rho_b}\right)
\left(\frac{\rho_b^{\rm pl}}{\rho_a^{\rm pl}}\right),
\label{Halldef_anisotropic}
\end{equation}
\noindent
where
$R_{\rm H}^{\rm pl}$ is
the Hall coefficient of the CuO$_2$ planes which can be anisotropic
and $\rho_{a(b)}^{\rm pl}$ is the resistivity
of the CuO$_2$ planes along the $a$($b$)-axis.
Unfortunately, $\rho_{b}^{\rm pl}$ is not measurable
and thus $R_{\rm H}^{\rm pl}$ cannot be obtained
if the CuO$_2$ planes are anisotropic and the Cu-O chains are conductive.

\newcommand{\FigureFive}{
\begin{figure}
\includegraphics[width=\figurewidthwide]{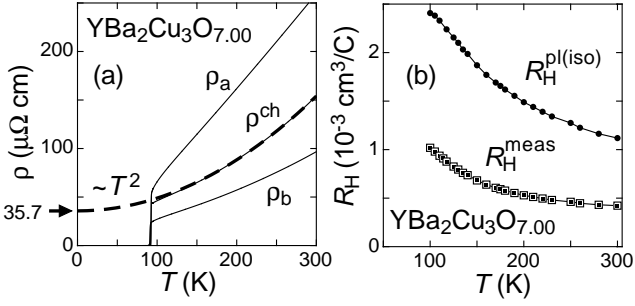}
\vspace{0.2cm}
\caption{(a) $T$ dependences of $\rho_a$, $\rho_b$ and $\rho^{\rm ch}$
for samples with $y=7.00$, where $\rho^{\rm ch}$ is obtained by using the formula
$1/\rho^{\rm ch}=1/\rho_b-1/\rho_a$. The dashed line is a fit to $\rho^{\rm ch}(T)$
by $aT^2+b$. (b) $T$ dependences of $R_{\rm H}^{\rm pl(iso)}$
and $R_{\rm H}^{\rm meas}$ with $y=7.00$.}
\label{Fig_y700}
\end{figure}}
\ifDC
\FigureFive
\fi

Figure \ref{Fig_y700}(b) shows the temperature dependence of
$R_{\rm H}^{\rm meas}$ and $R_{\rm H}^{\rm pl(iso)}$
with $y=7.00$.
$R_{\rm H}^{\rm pl(iso)}$ becomes more than two times larger than
$R_{\rm H}^{\rm meas}$ in the slightly overdoped samples.
Figure \ref{RH_plane_Figure}(a) shows the temperature dependence of
$R_{\rm H}^{\rm pl(iso)}$ for samples with $y\ge 6.60$.
The significant decrease at low temperatures in samples with $y\sim 6.75$
we noted in Fig. \ref{measured_RH_Figure} does not diminish
in $R_{\rm H}^{\rm pl(iso)}$.
Figures \ref{RH_plane_Figure}(b) and \ref{RH_plane_Figure}(c) show
the $y$-denpendences of $R_{\rm H}^{\rm meas}$ and
$R_{\rm H}^{\rm pl(iso)}$ at 150 K and 300 K, respectively.
One should note that
the obtained $R_{\rm H}^{\rm pl(iso)}$ just gives the maximum possible value
of $R_{\rm H}^{\rm pl}$
in the limit where the Cu-O chains are conductive and
the CuO$_2$ planes are isotropic.
On the other hand, if the Cu-O chains are insulating,
$R_{\rm H}^{\rm pl}$ becomes equal to
$R_{\rm H}^{\rm meas}$.
Therefore, the above analysis of the Hall coefficient
provides a range of the $R_{\rm H}^{\rm pl}$ values.
In the heavily underdoped samples where chains are broken,
the chain conduction is expected to be negligible and
the observed peculiar anisotropy
is most likely due to the planes \cite{AndoAnisotropy}; in this case, one can see from
Eq. (\ref{Halldef_anisotropic}) that $R_{\rm H}^{\rm meas}$ is
a true measure of \RH of the planes.

\newcommand{\FigureSix}{
\begin{figure}
\includegraphics[width=\figurewidthwide]{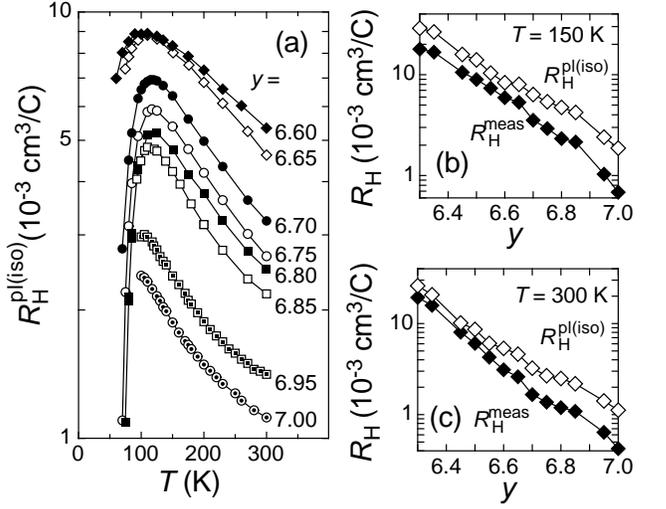}
\vspace{0.2cm}
\caption{(a) $T$ dependences of $R_{\rm H}^{\rm pl(iso)}$ for samples with $y\ge 6.60$.
(b, c) $y$ dependences of $R_{\rm H}^{\rm meas}$ and $R_{\rm H}^{\rm pl(iso)}$ at 150 and 300 K.}
\label{RH_plane_Figure}
\end{figure}}
\ifDC
\FigureSix
\fi

\subsection{Hall conductivity}

In contrast to the Hall resistivity (Hall coefficient),
we can unambiguously obtain the Hall {\it conductivity}
of the CuO$_2$ planes
regardless of anisotropy of the CuO$_2$ planes.
Since the Hall conductivity is calculated by
$\sigma_{xy} = \rho_{yx}/(\rho_{xx}\rho_{yy}-\rho_{yx}\rho_{yx})
\simeq\rho_{yx}/(\rho_{xx}\rho_{yy})$,
in YBCO we calculate it as
\begin{equation}
\sigma_{xy}^{\rm pl} =
\frac{\rho_{yx}^{\rm pl}}{\rho_{a}^{\rm pl}\rho_{b}^{\rm pl}},
\label{sigma_xy_def}
\end{equation}
\noindent
where $\sigma_{xy}^{\rm pl}$ is the Hall conductivity
of the CuO$_2$ planes and
$\rho_{yx}^{\rm pl}$ is the Hall resistivity of the CuO$_2$ planes.
We can transform Eq. (\ref{sigma_xy_def})
by using Eq. (\ref{Halldef_anisotropic}) into
\[
\sigma_{xy}^{\rm pl} =
\frac{\rho_{yx}^{\rm meas}}{\rho_{a}\rho_{b}}.
\]
\noindent
Note that this result tells us that the Hall conductivity in the CuO$_2$
planes is the same as that of the total system
regardless of the anisotropy of the CuO$_2$ planes,
which means that $\sigma_{xy}^{\rm pl} = \sigma_{xy}$ always holds
in YBCO. Thus, in the following
we do not discriminate $\sigma_{xy}^{\rm pl}$ from $\sigma_{xy}$.

\newcommand{\FigureSeven}{
\begin{figure}
\includegraphics[width=\figurewidthwide]{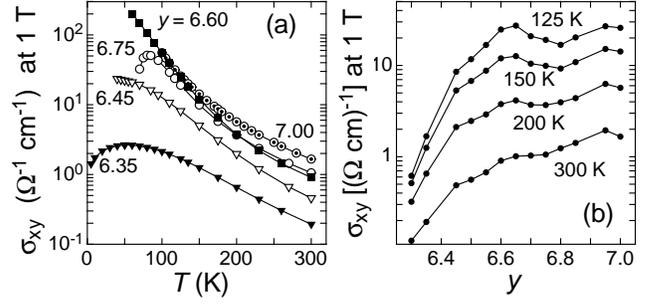}
\vspace{0.2cm}
\caption{(a) $T$ dependences of $\sigma_{xy}$ for
samples with 5 representative $y$ values. (b) $y$ dependences of $\sigma_{xy}$
at 125, 150, 200 and 300 K.}
\label{sigma_xy_figure}
\end{figure}}
\ifDC
\FigureSeven
\fi

Figure \ref{sigma_xy_figure}(a) shows the temperature dependences of $\sigma_{xy}$
for samples with 5 representative compositions in a semi-log plot.
We omitted other samples only for clarity.
For the samples with $y\sim$ 6.75 the low-temperature decrease
remains in $\sigma_{xy}$ and thus the observed decrease
in $R_{\rm H}$ at low temperatures
is not an artifact of some feature of the Cu-O chains
but is an intrinsic feature of the CuO$_2$ planes.
The $y$ dependence of $\sigma_{xy}$ at fixed temperatures
is shown in Fig. \ref{sigma_xy_figure}(b).
$\sigma_{xy}$ at low temperatures
appears to be suppressed at the oxygen content of $\sim6.75$
\citeSegawa;
this anomaly also has nothing to do with the Cu-O chains.
As we discussed in Ref. \citeSegawaAnother  , this anomaly
must be due to a peculiar doping dependence of the mobility of the carriers.

\subsection{Hall angle}
Next, let us analyze the cotangent of the Hall angle ($\cot \Theta_{\rm H}$).
As mentioned above,
the Hall coefficient $R_{\rm H}$ appears to show no in-plane anisotropy
as far as magnetic fields are applied along the $c$-axis,
which is expected from the Onsager relation.
On the other hand, the in-plane resistivity is observed to be anisotropic \cite{AndoAnisotropy}
in a whole doping range in YBCO.
The parameter $\cot \Theta_{\rm H}$ is calculated by $\rho_{xx}/\rho_{yx}$,
and thus in general $\cot \Theta_{\rm H}$ is expected to be anisotropic.
With careful considerations, it turns out that
one can obtain the Hall angle of the CuO$_2$ planes
from measurable properties only in the $\rho_b$-sample.
We can express $\cot \Theta_{\rm H}$
($\equiv E_x/E_y=\rho_{xx}/\rho_{yx}$)
of the CuO$_2$ planes along the $b$-axis as
\[
\cot \Theta_{{\rm H}(b)}^{\rm pl}=\frac{\rho_b^{\rm pl}}{\rho_{yx}^{\rm pl}},
\]
\noindent
where $\Theta_{{\rm H}(b)}^{\rm pl}$ is the Hall angle
of the CuO$_2$ planes with the current along the $b$-axis,
and from Eq. (\ref{Halldef_anisotropic}) we obtain
\[
\cot \Theta_{{\rm H}(b)}^{\rm pl} = \frac{\rho_b}{\rho_{yx}^{\rm meas}},
\]
\noindent
where the non-measurable property $\rho_b^{\rm pl}$ is canceled out.
%

\newcommand{\FigureEight}{
\begin{figure}
\includegraphics[width=\figurewidthwide]{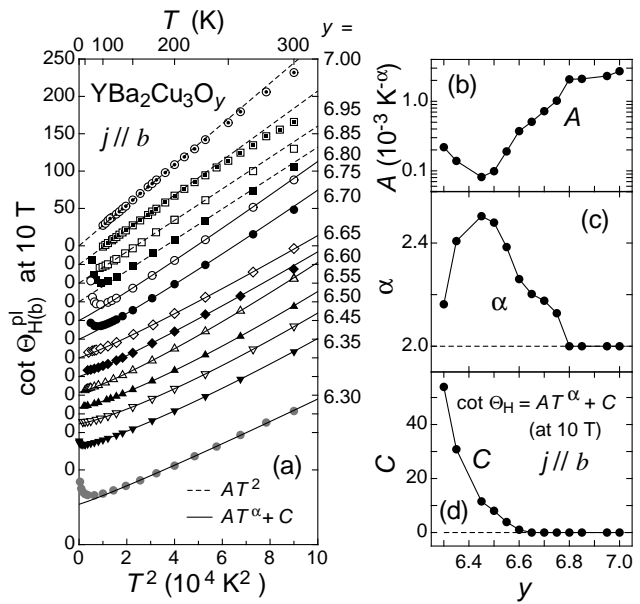}
\caption{(a) $\cot \Theta_{{\rm H}(b)}^{\rm pl}$ vs $T^2$ plot for $y=6.30$--7.00, where dashed lines are fits to the data by $AT^2$ and solid lines are fits to the data by $AT^\alpha+C$.
(b, c, d) $y$ dependences of the fitting parameters $A$, $\alpha$ and $C$, respectively.}
\label{cot_Figure_b}
\end{figure}}
\ifDC
\FigureEight
\fi

Fig. \ref{cot_Figure_b}(a) shows a
$\cot \Theta_{\rm H(b)}^{\rm pl}$ vs. $T^2$ plot
for samples with $y=6.30$--7.00,
where the origin for each sample is shifted for clarity.
In highly doped samples with $y = 6.95$ and 7.00
$\cot \Theta_{\rm H}$ appears to be fitted well
by a function of $AT^2$ except for high temperatures
(dashed lines in the figure).
Note that only one fitting parameter is used for the fitting
and the residual component is zero.
This result is consistent with the reported result of $\rho_b/\rho_{yx}$ for a 90-K sample
\cite{JPRice}.
The data for $y=$ 6.85 and 6.80 can also be fitted by the
function of $AT^2$ (dashed lines in the figure); however, 
at low temperatures some upward deviation is observed.
The low-temperature deviation seen in samples with $y= 6.60$--6.85
is clearly corresponding to the decrease in \RH at low tempeartures
and thus this anomaly seems to be one of the 60-K anomalies
rather than an effect of the pseudogap \cite{Xu_Preprint,AbeZnYBCO}.
The $T$-dependences of $\cot \Theta_{\rm H}$ for samples with $y\le 6.75$
can be fit by a function of $AT^\alpha +C$ with $\alpha >2$
(solid lines in the figure) except for the upturn at low temperatures.
Figures \ref{cot_Figure_b}(b-d) show $y$ dependences of
the fitting parameters.
The power $\alpha$ is 2 for $y \ge 6.80$ and
increases with decreasing oxygen content from 6.75.
In the slightly-doped region $\alpha(y)$ shows a peak at $y\sim 6.45$
and decreases with decreasing $y$;
$\alpha$ is shown to be 2 in the low doping limit \cite{AndoHall03}.
The `residual' component $C$ becomes finite only for $y$ below $\sim 6.60$,
and this is reminiscent of the result of the normal-state orbital
magnetoresistance \cite{YBCO_MR02},
in which the residual component $b$
in its temperature dependence $(aT^2+b)^{-2}$
becomes finite for $y\le 6.60$.

\newcommand{\FigureNine}{
\begin{figure}
\includegraphics[width=\figurewidthwide]{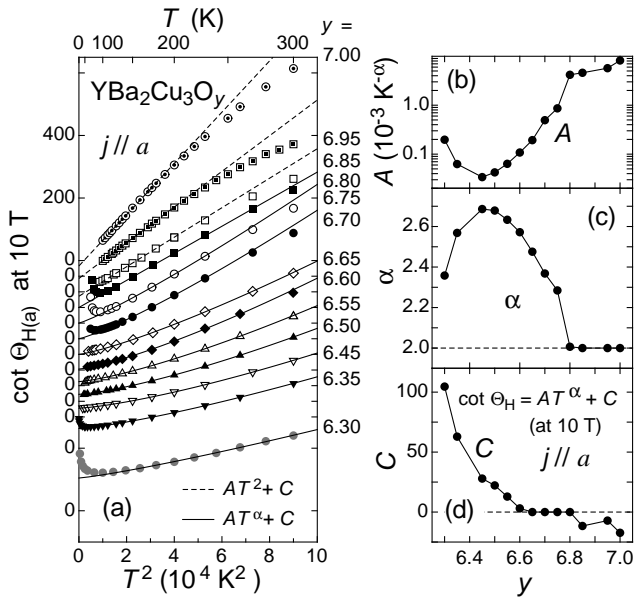}
\caption{(a) $\cot \Theta_{{\rm H}(a)}$ vs $T^2$ plot for $y=6.30$--7.00, where solid lines are fits to the data by $AT^\alpha+C$.
(b, c, d) $y$ dependences of the fitting parameters $A$, $\alpha$ and $C$, respectively.}
\label{cot_Figure_a}
\end{figure}}
\ifDC
\FigureNine
\fi

Of course we can also calculate $\cot \Theta_{\rm H}$ with the current along the $a$-axis
($\cot \Theta_{\rm H(a)}$),
which may contain contribution from the Cu-O chains
if the Cu-O chains are conductive.
Figure \ref{cot_Figure_a}(a) shows a $\cot \Theta_{\rm H(a)}$ vs. $T^2$
plot for samples with $y$=6.30--7.00
(the origin for each sample is shifted for clarity).
In highly doped samples with $y \ge 6.85$,
$\cot \Theta_{\rm H}$ appears to be fitted well
by a function of $AT^2+C$ except for high temperatures
(dashed lines in the figure);
however, unlike $\cot \Theta_{{\rm H}(b)}^{\rm pl}$,
the residual component $C$ becomes negative.
The solid lines are fits to the data by a function $AT^{\alpha}+C$
with $\alpha>2$.
For $y=6.70-6.80$ the data can be fitted only below $\sim$200 K,
whereas for more underdoped samples with $y\le 6.65$
the data are fitted well.
The $y$ dependences of the fitting parameters
are shown in Figs. \ref{cot_Figure_a}(b-d).
Observed tendency is qualitatively similar to that in $\cot \Theta_{{\rm H}(b)}^{\rm pl}$
except for the residual component for highly doped samples with $y\ge 6.85$.

From the Hall angle we can extract the Hall mobility $\mu_{\rm H}$;
Figs. \ref{Fig_muH}(a) and \ref{Fig_muH}(b) show the $y$ dependences of $\mu_{\rm H}$
with the current along the $b$-axis and the $a$-axis, respectively, at various temperatures.
Of course, the meaning of $\mu_{\rm H}$ with the current along the $b$-axis is more transparent,
but both data of $\mu_{\rm H}$ are qualitatively consistent.
At low temperatures the Hall mobility increases with decreasing
$y$ from 6.80 to 6.65.
This enhancement of the mobility compensates the effect
of decreasing carrier concentration and thus
an overlapping of $\rho_a(T)$ is observed for samples with $y=6.65$--6.80
\citeSegawa.

\newcommand{\FigureTen}{
\begin{figure}
\includegraphics[width=\figurewidthwide]{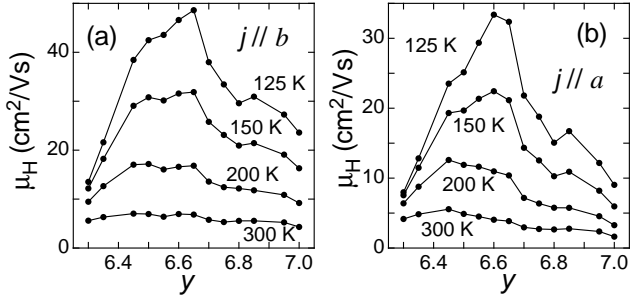}
\vspace{0.2cm}
\caption{$y$ dependences of the Hall mobility $\mu_{\rm H}$ at 125, 150, 200 and 300 K
with the current along the $b$-axis (a) and the $a$-axis (b).}
\label{Fig_muH}
\end{figure}}
\ifDC
\FigureTen
\fi

\subsection{Comparison with YBa$_2$Cu$_4$O$_8$}
YBa$_2$Cu$_4$O$_8$ (Y124) can be a good reference
system of YBCO, because Y124 has doubled Cu-O chains
which are believed to show good conduction.
Figures \ref{RH_Y124_Figure}(a-d) shows the temperature dependences of
$\rho_a$, $\rho_b$,  $R_{\rm H}^{\rm meas}$, $R_{\rm H}^{\rm pl(iso)}$,
$\sigma_{xy}$ and $\cot \Theta_{\rm H(b)}^{\rm pl}$
for our YBCO samples ($y=7.00$ and/or 6.75)
together with those for Y124,
which are extracted
from the data in published papers \cite{Bucher,HusseyPRB97}.
Fig. \ref{RH_Y124_Figure}(a) shows the temperature
dependences of $\rho_a$ and $\rho_b$.
While these data are not very different between YBCO$_{7.00}$ and Y124,
the anisotropy of the resistivity in Y124 is
larger than that in YBCO$_{7.00}$; this strong anisotropy is considered to
come from the doubled Cu-O chains.
The temperature dependence of the measured Hall coefficient
is very different [Fig. \ref{RH_Y124_Figure}(b)];
that of Y124 is much weaker than YBCO.
However, when we calculate $R_{\rm H}^{\rm pl(iso)}$ from the data,
it shows qualitatively
similar behavior in both YBCO$_{7.00}$ and Y124.
This fact suggests that the very conductive Cu-O chains
mask a strong temperature-dependence of the Hall coefficient in Y124.
Figure \ref{RH_Y124_Figure}(c) shows $\sigma_{xy}(T)$
in Y124 and YBCO samples with $y=7.00$ and 6.75.
The Hall conductivity of Y124 appears to be very close to
that of YBCO$_{6.75}$ rather than YBCO$_{7.00}$.
This is understandable because Y124 is naturally underdoped.
The cotangent of the Hall angle along the $b$-axis is shown
in Fig. \ref{RH_Y124_Figure}(d).
Data for Y124 can be fitted well by a function of $AT^2$ below $\sim$ 230 K (solid lines).
Thus, the transport properties of the CuO$_2$ planes in Y124,
which are extracted by the analysis presented in this paper,
appear to be similar to those in YBCO, and
this fact gives confidence in the validity of our analysis.

\newcommand{\FigureEleven}{
\begin{figure}
\includegraphics[width=\figurewidthwide]{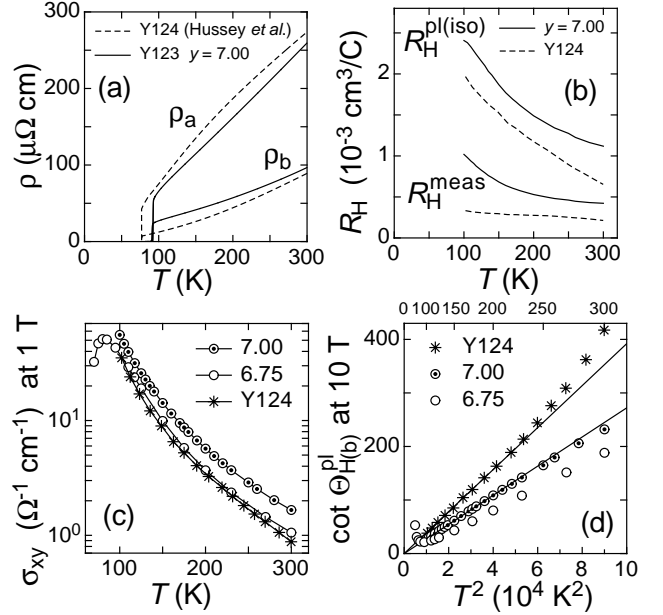}
\vspace{0.2cm}
\caption{The temperature dependences of (a) $\rho_a$ and $\rho_b$,
(b) $R_{\rm H}^{\rm meas}$ and $R_{\rm H}^{\rm pl(iso)}$
for Y124 and YBCO samples with $y=7.00$.
(c) $T$ dependences of $\sigma_{xy}$ and
(d) $\cot \Theta_{\rm H(b)}^{\rm pl}$ vs. $T^2$ plot
for Y124 and YBCO samples with $y=7.00$ and 6.75.}
\label{RH_Y124_Figure}
\end{figure}}
\ifDC
\FigureEleven
\fi

\subsection{Finite Hall coefficient in quasi-1D systems}
\label{quasi1D}

Our analysis assumes that the Cu-O chains do not contribute
to the Hall electric-field, whereas
in actual quasi-1D systems the Hall resistivity
has been observed to be not negligible
\cite{Horii02,Mihaly00}.
This is mostly due to the large resistivity in the direction
perpendicular to the chains, $\rho_{yy}$.
One can easily understand this
situation by noticing that the Hall resistivity $\rho_{yx}$ is expressed
as \[ \rho_{yx} = \sigma_{xy} \rho_{xx} \rho_{yy},\]
where $\sigma_{xy}$ is the Hall conductivity of the quasi-1D system and
$\rho_{xx}$ is the resistivity along the chain direction. (In quasi-1D
metals, $\rho_{yy} \gg \rho_{xx}$.) This formula tells us that a sizable
$\rho_{yx}$ can be observed for a very small $\sigma_{xy}$, if $\rho_{yy}$ is
very large. In fact, if we calculate $\sigma_{xy}$ for the case of
PrBa$_2$Cu$_4$O$_8$(Pr124) using published data \cite{Horii02},
it turns out that $\sigma_{xy}$ of
Pr124 is only 10$^{-4}$ of that of YBCO.
This estimate indicates that the contribution of $\sigma_{xy}$
of the Cu-O chains to the total
Hall conductivity is negligible in YBCO. More generally speaking, in a
chain-plane-composite system, the Hall conductivity of the total system,
$\sigma_{xy}^{\rm tot}$, is given by
\[
  \sigma_{xy}^{\rm tot} = \sigma_{xy}^{\rm pl} + \sigma_{xy}^{\rm ch},
\]
where $\sigma_{xy}^{\rm pl}$ and $\sigma_{xy}^{\rm ch}$
are the individual Hall conductivity of the planes and the chains. In the chain layer,
$\sigma_{xy}^{\rm ch}$ is always very small
even when $\rho_{yx}^{\rm ch}$ is sizable,
as we discussed above. Therefore, one can safely neglect the
contribution of $\sigma_{xy}^{\rm ch}$ in the composite system and consider
that the behavior of $\sigma_{xy}^{\rm tot}$ is governed almost solely by
$\sigma_{xy}^{\rm pl}$.

\subsection{Anomalous decrease in \RH at low tempeartures}

Finally, let us discuss the pronounced decrease in $R_{\rm H}$ observed
only for samples near $y\simeq 6.75$ at low temperatures.
We emphasize that the present data are measured only at $T>T_{\rm c}$
by sweeping the magnetic fields up to at least 10 T,
where the field dependence of the Hall resistivity is
$H$-linear in the whole region.
Therefore, the decrease in \RH reflects the normal-state property
and is irrelevant
to the negative Hall anomaly
\cite{NegativeHall}
which is observed only at low fields and below $T_{\rm c}$.
Our observation is somewhat reminiscent
of the decrease in \RH observed in La$_{2-x-y}$Nd$_y$Sr$_x$CuO$_4$ \citeNoda,
and thus the present result may actually be related to charged stripes.
In fact, the carrier concentration of YBCO$_{6.75}$ is consistent with
the hole concentration $p\sim 1/8$ (per Cu atom in the CuO$_2$ planes)
\citeMOSETC,
where so-called `1/8 anomaly' is observed in cuprate superconductors;
therefore, it is likely that the observed anomaly in \RH is
related to the `1/8 problem'.
Note, however, that the in-plane resistivity anisotropy is observed to become {\it small}
in samples with $y\sim 6.75$ at low temperatures \cite{AndoAnisotropy}.
Such a result may seem to imply that
one-dimensionality of charge dynamics in the CuO$_2$ planes is minimal,
which is inconsistent with the picture of static stripes at $p\sim 1/8$.
However, the possibility of fluctuating charge stripes is not excluded,
because the stripe liquid can be in an `isotropic' phase \cite{Kivelson98};
in this case,
the decrease in \RH  can be caused
by a realization of the particle-hole symmetry
when the hole concentration becomes $\sim 1/8$ and
the stripes become 1/4 filled \cite{EmeryPRL00,Prelovsek01}.

\section{Summary}
In summary,
we show how to extract the Hall conductivity and the cotangent of the Hall angle
of the CuO$_2$ planes in YBCO
regardless of conduction of the Cu-O chains nor
in-plane anisotropy of the CuO$_2$ planes.
The analysis is applied to the data of untwinned YBCO for a wide range of doping
as well as the available data of Y124 in the literature.
The present analysis provides a legitimate way to discuss
the Hall effect of the CuO$_2$ planes in YBCO
on the same ground as that in other high-$T_{\rm c}$ cuprates. 

\section{Acknowledgments}

We would like to acknowledge A.N. Lavrov, N. P. Ong,
and S. Uchida for fruitful discussions and Y. Abe for technical
assistances.

\ifTEST
\section{Footnote}
For estimating the distribution of the current,
it is necessary to map an anisotropic sample to
an equivalent isotropic sample [L. J. van der Pauw, Philips Res. Repts. {\bf 16}, 187 (1961).].
In YBCO $\rho_b$ is always smaller than $\rho_a$ and thus
the $\rho_b$-sample is effectively shorter and wider
than the $\rho_a$-sample.
\RH becomes smaller when electrodes get closer to a current contact of a sample,
because the current contact is made on the whole surface
on which the Hall voltage is short-circuited.
Since in the $\rho_b$-sample electrodes become effectively closer to the
current contacts than in the $\rho_a$-sample,
$R_{\rm H}^b$ becomes slightly smaller than $R_{\rm H}^a$.
\fi

\medskip
\vfil

\ifTEST
\bibliography{segawa9a}

\begin{thebibliography}{39}
\expandafter\ifx\csname natexlab\endcsname\relax\def\natexlab#1{#1}\fi
\expandafter\ifx\csname bibnamefont\endcsname\relax
  \def\bibnamefont#1{#1}\fi
\expandafter\ifx\csname bibfnamefont\endcsname\relax
  \def\bibfnamefont#1{#1}\fi
\expandafter\ifx\csname citenamefont\endcsname\relax
  \def\citenamefont#1{#1}\fi
\expandafter\ifx\csname url\endcsname\relax
  \def\url#1{\texttt{#1}}\fi
\expandafter\ifx\csname urlprefix\endcsname\relax\def\urlprefix{URL }\fi
\providecommand{\bibinfo}[2]{#2}
\providecommand{\eprint}[2][]{\url{#2}}

\bibitem[{\citenamefont{Ong}()}]{OngReview}
N. P. Ong, in {\it Physical Properties of High Temperature Superconductors},
edited by D. M. Ginsberg (World Scientific, Singapore, 1990), Vol. 2, p. 459.

\bibitem[{\citenamefont{Chien et~al.}(1991)\citenamefont{Chien, Wang, and
  Ong}}]{Chien91}
\bibinfo{author}{\bibfnamefont{T.~R.} \bibnamefont{Chien}},
  \bibinfo{author}{\bibfnamefont{Z.~Z.} \bibnamefont{Wang}}, \bibnamefont{and}
  \bibinfo{author}{\bibfnamefont{N.~P.} \bibnamefont{Ong}},
  \bibinfo{journal}{Phys.\ Rev.\ Lett.} \textbf{\bibinfo{volume}{67}},
  \bibinfo{pages}{2088} (\bibinfo{year}{1991}).

\bibitem[{\citenamefont{Anderson}(1991)}]{Anderson91}
\bibinfo{author}{\bibfnamefont{P.~W.} \bibnamefont{Anderson}},
  \bibinfo{journal}{Phys.\ Rev.\ Lett.} \textbf{\bibinfo{volume}{67}},
  \bibinfo{pages}{2092} (\bibinfo{year}{1991}).

\bibitem[{\citenamefont{Carrington et~al.}(1992)\citenamefont{Carrington,
  Mackenzie, Lin, and Cooper}}]{Carrington92}
\bibinfo{author}{\bibfnamefont{A.}~\bibnamefont{Carrington}},
  \bibinfo{author}{\bibfnamefont{A.~P.} \bibnamefont{Mackenzie}},
  \bibinfo{author}{\bibfnamefont{C.~T.} \bibnamefont{Lin}}, \bibnamefont{and}
  \bibinfo{author}{\bibfnamefont{J.~R.} \bibnamefont{Cooper}},
  \bibinfo{journal}{Phys.\ Rev.\ Lett.} \textbf{\bibinfo{volume}{69}},
  \bibinfo{pages}{2855} (\bibinfo{year}{1992}).

\bibitem[{\citenamefont{Alexandrov et~al.}(1994)\citenamefont{Alexandrov,
  Bratkovsky, and Mott}}]{Alexandrov94}
\bibinfo{author}{\bibfnamefont{A.~S.} \bibnamefont{Alexandrov}},
  \bibinfo{author}{\bibfnamefont{A.~M.} \bibnamefont{Bratkovsky}},
  \bibnamefont{and} \bibinfo{author}{\bibfnamefont{N.~F.} \bibnamefont{Mott}},
  \bibinfo{journal}{Phys.\ Rev.\ Lett.} \textbf{\bibinfo{volume}{72}},
  \bibinfo{pages}{1734} (\bibinfo{year}{1994}).

\bibitem[{\citenamefont{Kotliar and Varma}(1996)}]{Kotliar96}
\bibinfo{author}{\bibfnamefont{G.}~\bibnamefont{Kotliar}},
  \bibinfo{author}{\bibfnamefont{A.}~\bibnamefont{Sengupta}}, \bibnamefont{and}
  \bibinfo{author}{\bibfnamefont{C.~M.} \bibnamefont{Varma}},
  \bibinfo{journal}{Phys.\ Rev.\ B} \textbf{\bibinfo{volume}{53}},
  \bibinfo{pages}{3573} (\bibinfo{year}{1996}).

\bibitem[{\citenamefont{Coleman et~al.}(1996)\citenamefont{Coleman, Schofield,
  and Tsvelik}}]{Coleman96}
\bibinfo{author}{\bibfnamefont{P.}~\bibnamefont{Coleman}},
  \bibinfo{author}{\bibfnamefont{A.~J.} \bibnamefont{Schofield}},
  \bibnamefont{and} \bibinfo{author}{\bibfnamefont{A.~M.}
  \bibnamefont{Tsvelik}}, \bibinfo{journal}{Phys.\ Rev.\ Lett.}
  \textbf{\bibinfo{volume}{76}}, \bibinfo{pages}{1324} (\bibinfo{year}{1996}).

\bibitem[{\citenamefont{Stojkovi\'{c} and Pines}(1997)}]{Stojkovic97}
\bibinfo{author}{\bibfnamefont{B.~P.} \bibnamefont{Stojkovi\'{c}}}
  \bibnamefont{and} \bibinfo{author}{\bibfnamefont{D.}~\bibnamefont{Pines}},
  \bibinfo{journal}{Phys.\ Rev.\ B} \textbf{\bibinfo{volume}{55}},
  \bibinfo{pages}{8576} (\bibinfo{year}{1997}).

\bibitem[{\citenamefont{Ioffe and Millis}(1998)}]{Ioffe98}
\bibinfo{author}{\bibfnamefont{L.~B.} \bibnamefont{Ioffe}} \bibnamefont{and}
  \bibinfo{author}{\bibfnamefont{A.~J.} \bibnamefont{Millis}},
  \bibinfo{journal}{Phys.\ Rev.\ B} \textbf{\bibinfo{volume}{58}},
  \bibinfo{pages}{11631} (\bibinfo{year}{1998}).

\bibitem[{\citenamefont{Zheleznyak et~al.}(1999)\citenamefont{Zheleznyak,
  Yakovenko, and Drew}}]{Zheleznyak99}
\bibinfo{author}{\bibfnamefont{A.~T.} \bibnamefont{Zheleznyak}},
  \bibinfo{author}{\bibfnamefont{V.~M.} \bibnamefont{Yakovenko}},
  \bibnamefont{and} \bibinfo{author}{\bibfnamefont{H.~D.} \bibnamefont{Drew}},
  \bibinfo{journal}{Phys.\ Rev.\ B} \textbf{\bibinfo{volume}{59}},
  \bibinfo{pages}{207} (\bibinfo{year}{1999}).

\bibitem[{\citenamefont{Kontani et~al.}(1999)\citenamefont{Kontani, Kanki, and
  Ueda}}]{Kontani99}
\bibinfo{author}{\bibfnamefont{H.}~\bibnamefont{Kontani}},
  \bibinfo{author}{\bibfnamefont{K.}~\bibnamefont{Kanki}}, \bibnamefont{and}
  \bibinfo{author}{\bibfnamefont{K.}~\bibnamefont{Ueda}},
  \bibinfo{journal}{Phys.\ Rev.\ B} \textbf{\bibinfo{volume}{59}},
  \bibinfo{pages}{14723} (\bibinfo{year}{1999}).

\bibitem[{\citenamefont{Varma and Abrahams}(2001)}]{Varma01}
\bibinfo{author}{\bibfnamefont{C.~M.} \bibnamefont{Varma}} \bibnamefont{and}
  \bibinfo{author}{\bibfnamefont{E.}~\bibnamefont{Abrahams}},
  \bibinfo{journal}{Phys.\ Rev.\ Lett.} \textbf{\bibinfo{volume}{86}},
  \bibinfo{pages}{4652} (\bibinfo{year}{2001}).

\bibitem[{\citenamefont{Gagnon et~al.}(1994)\citenamefont{Gagnon, Lupien, and
  Taillefer}}]{Gagnon}
\bibinfo{author}{\bibfnamefont{R.}~\bibnamefont{Gagnon}},
  \bibinfo{author}{\bibfnamefont{C.}~\bibnamefont{Lupien}}, \bibnamefont{and}
  \bibinfo{author}{\bibfnamefont{L.}~\bibnamefont{Taillefer}},
  \bibinfo{journal}{Phys.\ Rev.\ B} \textbf{\bibinfo{volume}{50}},
  \bibinfo{pages}{3458} (\bibinfo{year}{1994}).

\bibitem[{\citenamefont{Takenaka et~al.}(1994)\citenamefont{Takenaka,
  Mizuhashi, Takagi, and Uchida}}]{Takenaka}
\bibinfo{author}{\bibfnamefont{K.}~\bibnamefont{Takenaka}},
  \bibinfo{author}{\bibfnamefont{K.}~\bibnamefont{Mizuhashi}},
  \bibinfo{author}{\bibfnamefont{H.}~\bibnamefont{Takagi}}, \bibnamefont{and}
  \bibinfo{author}{\bibfnamefont{S.}~\bibnamefont{Uchida}},
  \bibinfo{journal}{Phys.\ Rev.\ B} \textbf{\bibinfo{volume}{50}},
  \bibinfo{pages}{R6534} (\bibinfo{year}{1994}).

\bibitem[{\citenamefont{Ando et~al.}(2002)\citenamefont{Ando, Segawa, Komiya,
  and Lavrov}}]{AndoAnisotropy}
\bibinfo{author}{\bibfnamefont{Y.}~\bibnamefont{Ando}},
  \bibinfo{author}{\bibfnamefont{K.}~\bibnamefont{Segawa}},
  \bibinfo{author}{\bibfnamefont{S.}~\bibnamefont{Komiya}}, \bibnamefont{and}
  \bibinfo{author}{\bibfnamefont{A.~N.} \bibnamefont{Lavrov}},
  \bibinfo{journal}{Phys.\ Rev.\ Lett.} \textbf{\bibinfo{volume}{88}},
  \bibinfo{pages}{137005} (\bibinfo{year}{2002}).

\bibitem[{\citenamefont{Hoffmann et~al.}(1993)\citenamefont{Hoffmann, Manuel,
  Peter, Walker, Gauthier, Shukla, Barbiellini, Massidda, Adam, Hardy
  et~al.}}]{Hoffmann}
\bibinfo{author}{\bibfnamefont{L.}~\bibnamefont{Hoffmann}},
  \bibinfo{author}{\bibfnamefont{A.~A.} \bibnamefont{Manuel}},
  \bibinfo{author}{\bibfnamefont{M.}~\bibnamefont{Peter}},
  \bibinfo{author}{\bibfnamefont{E.}~\bibnamefont{Walker}},
  \bibinfo{author}{\bibfnamefont{M.}~\bibnamefont{Gauthier}},
  \bibinfo{author}{\bibfnamefont{A.}~\bibnamefont{Shukla}},
  \bibinfo{author}{\bibfnamefont{B.}~\bibnamefont{Barbiellini}},
  \bibinfo{author}{\bibfnamefont{S.}~\bibnamefont{Massidda}},
  \bibinfo{author}{\bibfnamefont{Gh.}~\bibnamefont{Adam}},
  \bibinfo{author}{\bibfnamefont{S.}~\bibnamefont{Adam}},
  \bibinfo{author}{\bibfnamefont{W.~N.} \bibnamefont{Hardy}}, \bibnamefont{and}
  \bibinfo{author}{\bibfnamefont{R.}~\bibnamefont{Liang}},
  \bibinfo{journal}{Phys.\ Rev.\ Lett.}
  \textbf{\bibinfo{volume}{71}}, \bibinfo{pages}{4047} (\bibinfo{year}{1993}).

\bibitem[{\citenamefont{Rice et~al.}(1991)\citenamefont{Rice, Giapintzakis,
  Ginsberg, and Mochel}}]{JPRice}
\bibinfo{author}{\bibfnamefont{J.~P.} \bibnamefont{Rice}},
  \bibinfo{author}{\bibfnamefont{J.}~\bibnamefont{Giapintzakis}},
  \bibinfo{author}{\bibfnamefont{D.~M.} \bibnamefont{Ginsberg}},
  \bibnamefont{and} \bibinfo{author}{\bibfnamefont{J.~M.}
  \bibnamefont{Mochel}}, \bibinfo{journal}{Phys.\ Rev.\ B}
  \textbf{\bibinfo{volume}{44}}, \bibinfo{pages}{10158} (\bibinfo{year}{1991}).

\bibitem[{\citenamefont{Segawa and Ando}(1999)}]{SegawaZndoped}
\bibinfo{author}{\bibfnamefont{K.}~\bibnamefont{Segawa}} \bibnamefont{and}
  \bibinfo{author}{\bibfnamefont{Y.}~\bibnamefont{Ando}},
  \bibinfo{journal}{Phys.\ Rev.\ B} \textbf{\bibinfo{volume}{59}},
  \bibinfo{pages}{R3948} (\bibinfo{year}{1999}).

\bibitem[{\citenamefont{Lavrov and Kozeeva}(1995)}]{Lavrov}
\bibinfo{author}{\bibfnamefont{A.}~\bibnamefont{Lavrov}} \bibnamefont{and}
  \bibinfo{author}{\bibfnamefont{L.}~\bibnamefont{Kozeeva}},
  \bibinfo{journal}{Physica C} \textbf{\bibinfo{volume}{253C}},
  \bibinfo{pages}{313} (\bibinfo{year}{1995}), and references therein.

\bibitem[{\citenamefont{Segawa and Ando}(2001)}]{SegawaOverlap}
\bibinfo{author}{\bibfnamefont{K.}~\bibnamefont{Segawa}} \bibnamefont{and}
  \bibinfo{author}{\bibfnamefont{Y.}~\bibnamefont{Ando}},
  \bibinfo{journal}{Phys.\ Rev.\ Lett.} \textbf{\bibinfo{volume}{86}},
  \bibinfo{pages}{4907} (\bibinfo{year}{2001}).

\bibitem[{\citenamefont{Kishio et~al.}(1987)\citenamefont{Kishio, Shimoyama,
  Hasegawa, Kitazawa, and Fueki}}]{Kishio}
\bibinfo{author}{\bibfnamefont{K.}~\bibnamefont{Kishio}},
  \bibinfo{author}{\bibfnamefont{J.}~\bibnamefont{Shimoyama}},
  \bibinfo{author}{\bibfnamefont{T.}~\bibnamefont{Hasegawa}},
  \bibinfo{author}{\bibfnamefont{K.}~\bibnamefont{Kitazawa}}, \bibnamefont{and}
  \bibinfo{author}{\bibfnamefont{K.}~\bibnamefont{Fueki}},
  \bibinfo{journal}{Jpn. J. Appl. Phys.} \textbf{\bibinfo{volume}{26}},
  \bibinfo{pages}{L1228} (\bibinfo{year}{1987}).

\bibitem[{\citenamefont{Abe et~al.}(1999)\citenamefont{Abe, Segawa, and
  Ando}}]{AbeZnYBCO}
\bibinfo{author}{\bibfnamefont{Y.}~\bibnamefont{Abe}},
  \bibinfo{author}{\bibfnamefont{K.}~\bibnamefont{Segawa}}, \bibnamefont{and}
  \bibinfo{author}{\bibfnamefont{Y.}~\bibnamefont{Ando}},
  \bibinfo{journal}{Phys.\ Rev.\ B} \textbf{\bibinfo{volume}{60}},
  \bibinfo{pages}{R15055} (\bibinfo{year}{1999}).

\bibitem[{\citenamefont{Onsager}(1931)}]{Onsager}
\bibinfo{author}{\bibfnamefont{L.}~\bibnamefont{Onsager}},
  \bibinfo{journal}{Phys.\ Rev.} \textbf{\bibinfo{volume}{38}},
  \bibinfo{pages}{2265} (\bibinfo{year}{1931}).

\bibitem[{\citenamefont{Harris et~al.}(1994)\citenamefont{Harris, Yan, Tsui,
  Matsuda, and Ong}}]{Harris94}
\bibinfo{author}{\bibfnamefont{J.~M.} \bibnamefont{Harris}},
  \bibinfo{author}{\bibfnamefont{Y.~F.} \bibnamefont{Yan}},
  \bibinfo{author}{\bibfnamefont{O.~K.~C.} \bibnamefont{Tsui}},
  \bibinfo{author}{\bibfnamefont{Y.}~\bibnamefont{Matsuda}}, \bibnamefont{and}
  \bibinfo{author}{\bibfnamefont{N.~P.} \bibnamefont{Ong}},
  \bibinfo{journal}{Phys.\ Rev.\ Lett.} \textbf{\bibinfo{volume}{73}},
  \bibinfo{pages}{1711} (\bibinfo{year}{1994}).

\bibitem[{Not()}]{Note}
For estimating the distribution of the current,
it is necessary to map an anisotropic sample to
an equivalent isotropic sample [L. J. van der Pauw, Philips Res. Repts. {\bf 16}, 187 (1961).].
In YBCO $\rho_b$ is always smaller than $\rho_a$ and thus
the $\rho_b$-sample is effectively shorter and wider
than the $\rho_a$-sample.
\RH becomes smaller when electrodes get closer to a current contact of a sample,
because the current contact is made on the whole surface
on which the Hall voltage is short-circuited.
Since in the $\rho_b$-sample electrodes become effectively closer to the
current contacts than in the $\rho_a$-sample,
$R_{\rm H}^b$ becomes slightly smaller than $R_{\rm H}^a$.

\bibitem[{\citenamefont{Ito et~al.}(1993)\citenamefont{Ito, Takenaka, and
  Uchida}}]{Ito}
\bibinfo{author}{\bibfnamefont{T.}~\bibnamefont{Ito}},
  \bibinfo{author}{\bibfnamefont{K.}~\bibnamefont{Takenaka}}, \bibnamefont{and}
  \bibinfo{author}{\bibfnamefont{S.}~\bibnamefont{Uchida}},
  \bibinfo{journal}{Phys.\ Rev.\ Lett.} \textbf{\bibinfo{volume}{70}},
  \bibinfo{pages}{3995} (\bibinfo{year}{1993}).

\bibitem[{\citenamefont{Ando et~al.}(2000)\citenamefont{Ando, Hanaki, Ono,
  Murayama, Segawa, Miyamoto, and Komiya}}]{AndoCarrierConc}
\bibinfo{author}{\bibfnamefont{Y.}~\bibnamefont{Ando}},
  \bibinfo{author}{\bibfnamefont{Y.}~\bibnamefont{Hanaki}},
  \bibinfo{author}{\bibfnamefont{S.}~\bibnamefont{Ono}},
  \bibinfo{author}{\bibfnamefont{T.}~\bibnamefont{Murayama}},
  \bibinfo{author}{\bibfnamefont{K.}~\bibnamefont{Segawa}},
  \bibinfo{author}{\bibfnamefont{N.}~\bibnamefont{Miyamoto}}, \bibnamefont{and}
  \bibinfo{author}{\bibfnamefont{S.}~\bibnamefont{Komiya}},
  \bibinfo{journal}{Phys.\ Rev.\ B} \textbf{\bibinfo{volume}{61}},
  \bibinfo{pages}{R14956} (\bibinfo{year}{2000}).

\bibitem[{\citenamefont{Ando et~al.}(submitted)\citenamefont{Ando, Kurita,
  Komiya, Ono, and Segawa}}]{AndoHall03}
\bibinfo{author}{\bibfnamefont{Y.}~\bibnamefont{Ando}},
  \bibinfo{author}{\bibfnamefont{Y.}~\bibnamefont{Kurita}},
  \bibinfo{author}{\bibfnamefont{S.}~\bibnamefont{Komiya}},
  \bibinfo{author}{\bibfnamefont{S.}~\bibnamefont{Ono}}, \bibnamefont{and}
  \bibinfo{author}{\bibfnamefont{K.}~\bibnamefont{Segawa}}
  (\bibinfo{year}{submitted}).

\bibitem[{\citenamefont{Hussey et~al.}(1997)\citenamefont{Hussey, Nozawa,
  Takagi, Adachi, and Tanabe}}]{HusseyPRB97}
\bibinfo{author}{\bibfnamefont{N.~E.} \bibnamefont{Hussey}},
  \bibinfo{author}{\bibfnamefont{K.}~\bibnamefont{Nozawa}},
  \bibinfo{author}{\bibfnamefont{H.}~\bibnamefont{Takagi}},
  \bibinfo{author}{\bibfnamefont{S.}~\bibnamefont{Adachi}}, \bibnamefont{and}
  \bibinfo{author}{\bibfnamefont{K.}~\bibnamefont{Tanabe}},
  \bibinfo{journal}{Phys.\ Rev.\ B} \textbf{\bibinfo{volume}{56}},
  \bibinfo{pages}{R11423} (\bibinfo{year}{1997}).

\bibitem[{\citenamefont{Mih\'{a}ly et~al.}(2000)\citenamefont{Mih\'{a}ly,
  K\'{e}zsm\'{a}rki, Z\'{a}mborszky, and Forr\'{o}}}]{Mihaly00}
\bibinfo{author}{\bibfnamefont{G.}~\bibnamefont{Mih\'{a}ly}},
  \bibinfo{author}{\bibfnamefont{I.}~\bibnamefont{K\'{e}zsm\'{a}rki}},
  \bibinfo{author}{\bibfnamefont{F.}~\bibnamefont{Z\'{a}mborszky}},
  \bibnamefont{and}
  \bibinfo{author}{\bibfnamefont{L.}~\bibnamefont{Forr\'{o}}},
  \bibinfo{journal}{Phys.\ Rev.\ Lett.} \textbf{\bibinfo{volume}{84}},
  \bibinfo{pages}{2670} (\bibinfo{year}{2000});
  \bibinfo{author}{\bibfnamefont{J.}~\bibnamefont{Moser}},
  \bibinfo{author}{\bibfnamefont{J.~R.} \bibnamefont{Cooper}},
  \bibinfo{author}{\bibfnamefont{D.}~\bibnamefont{J\'{e}rome}},
  \bibinfo{author}{\bibfnamefont{B.}~\bibnamefont{Alavi}},
  \bibinfo{author}{\bibfnamefont{S.~E.} \bibnamefont{Brown}}, \bibnamefont{and}
  \bibinfo{author}{\bibfnamefont{K.}~\bibnamefont{Bechgaard}},
  \bibinfo{journal}{{\it ibid.}} \textbf{\bibinfo{volume}{84}},
  \bibinfo{pages}{2674} (\bibinfo{year}{2000}).
%

\bibitem[{\citenamefont{Horii et~al.}(2002)\citenamefont{Horii, Mizutani,
  Ikuta, Yamada, Ye, Matsushita, Hussey, Takagi, and Hirabayashi}}]{Horii02}
\bibinfo{author}{\bibfnamefont{S.}~\bibnamefont{Horii}},
  \bibinfo{author}{\bibfnamefont{U.}~\bibnamefont{Mizutani}},
  \bibinfo{author}{\bibfnamefont{H.}~\bibnamefont{Ikuta}},
  \bibinfo{author}{\bibfnamefont{Y.}~\bibnamefont{Yamada}},
  \bibinfo{author}{\bibfnamefont{J.~H.} \bibnamefont{Ye}},
  \bibinfo{author}{\bibfnamefont{A.}~\bibnamefont{Matsushita}},
  \bibinfo{author}{\bibfnamefont{N.~E.} \bibnamefont{Hussey}},
  \bibinfo{author}{\bibfnamefont{H.}~\bibnamefont{Takagi}}, \bibnamefont{and}
  \bibinfo{author}{\bibfnamefont{I.}~\bibnamefont{Hirabayashi}},
  \bibinfo{journal}{Phys.\ Rev.\ B} \textbf{\bibinfo{volume}{66}},
  \bibinfo{pages}{054530} (\bibinfo{year}{2002}).

\bibitem[{\citenamefont{Xu et~al.}(unpublished)\citenamefont{Xu, Zhang, and
  Ong}}]{Xu_Preprint}
\bibinfo{author}{\bibfnamefont{Z.~A.} \bibnamefont{Xu}},
  \bibinfo{author}{\bibfnamefont{Y.}~\bibnamefont{Zhang}}, \bibnamefont{and}
  \bibinfo{author}{\bibfnamefont{N.~P.} \bibnamefont{Ong}},
  \bibinfo{journal}{cond-mat/9903123}  (\bibinfo{year}{unpublished}).

\bibitem[{\citenamefont{Ando and Segawa}(2002{\natexlab{a}})}]{YBCO_MR02}
\bibinfo{author}{\bibfnamefont{Y.}~\bibnamefont{Ando}} \bibnamefont{and}
  \bibinfo{author}{\bibfnamefont{K.}~\bibnamefont{Segawa}},
  \bibinfo{journal}{Phys.\ Rev.\ Lett.} \textbf{\bibinfo{volume}{88}},
  \bibinfo{pages}{167005} (\bibinfo{year}{2002}{\natexlab{a}}).

\bibitem[{\citenamefont{Bucher et~al.}(1993)\citenamefont{Bucher, Steiner,
  Karpinski, Kaldis, and Wachter}}]{Bucher}
\bibinfo{author}{\bibfnamefont{B.}~\bibnamefont{Bucher}},
  \bibinfo{author}{\bibfnamefont{P.}~\bibnamefont{Steiner}},
  \bibinfo{author}{\bibfnamefont{J.}~\bibnamefont{Karpinski}},
  \bibinfo{author}{\bibfnamefont{E.}~\bibnamefont{Kaldis}}, \bibnamefont{and}
  \bibinfo{author}{\bibfnamefont{P.}~\bibnamefont{Wachter}},
  \bibinfo{journal}{Phys.\ Rev.\ Lett.} \textbf{\bibinfo{volume}{70}},
  \bibinfo{pages}{2012} (\bibinfo{year}{1993}).

\bibitem[{\citenamefont{Nagaoka et~al.}(1998)\citenamefont{Nagaoka, Matsuda,
  Obara, Sawa, Terashima, Chong, Takano, and Suzuki}}]{NegativeHall}
\bibinfo{author}{\bibfnamefont{T.}~\bibnamefont{Nagaoka}},
  \bibinfo{author}{\bibfnamefont{Y.}~\bibnamefont{Matsuda}},
  \bibinfo{author}{\bibfnamefont{H.}~\bibnamefont{Obara}},
  \bibinfo{author}{\bibfnamefont{A.}~\bibnamefont{Sawa}},
  \bibinfo{author}{\bibfnamefont{T.}~\bibnamefont{Terashima}},
  \bibinfo{author}{\bibfnamefont{I.}~\bibnamefont{Chong}},
  \bibinfo{author}{\bibfnamefont{M.}~\bibnamefont{Takano}}, \bibnamefont{and}
  \bibinfo{author}{\bibfnamefont{M.}~\bibnamefont{Suzuki}},
  \bibinfo{journal}{Phys.\ Rev.\ Lett.} \textbf{\bibinfo{volume}{80}},
  \bibinfo{pages}{3594} (\bibinfo{year}{1998}), and references therein.

\bibitem[{\citenamefont{Noda et~al.}(1999)\citenamefont{Noda, Eisaki, and
  Uchida}}]{Noda}
\bibinfo{author}{\bibfnamefont{T.}~\bibnamefont{Noda}},
  \bibinfo{author}{\bibfnamefont{H.}~\bibnamefont{Eisaki}}, \bibnamefont{and}
  \bibinfo{author}{\bibfnamefont{S.}~\bibnamefont{Uchida}},
  \bibinfo{journal}{Science} \textbf{\bibinfo{volume}{286}},
  \bibinfo{pages}{265} (\bibinfo{year}{1999}).

\bibitem[{\citenamefont{Ando and Segawa}(2002{\natexlab{b}})}]{AndoSNS}
\bibinfo{author}{\bibfnamefont{Y.}~\bibnamefont{Ando}} \bibnamefont{and}
  \bibinfo{author}{\bibfnamefont{K.}~\bibnamefont{Segawa}},
  \bibinfo{journal}{J. Phys. Chem. Solids} \textbf{\bibinfo{volume}{63}},
  \bibinfo{pages}{2253} (\bibinfo{year}{2002}{\natexlab{b}}).

\bibitem[{\citenamefont{Segawa and Ando}(2003)}]{SegawaMOS02}
\bibinfo{author}{\bibfnamefont{K.}~\bibnamefont{Segawa}} \bibnamefont{and}
  \bibinfo{author}{\bibfnamefont{Y.}~\bibnamefont{Ando}}, \bibinfo{journal}{J.
  Low Temp. Phys.} \textbf{\bibinfo{volume}{131}}, \bibinfo{pages}{821}
  (\bibinfo{year}{2003}).

\bibitem[{\citenamefont{Kivelson et~al.}(1998)\citenamefont{Kivelson, Fradkin,
  and Emery}}]{Kivelson98}
\bibinfo{author}{\bibfnamefont{S.~A.} \bibnamefont{Kivelson}},
  \bibinfo{author}{\bibfnamefont{E.}~\bibnamefont{Fradkin}}, \bibnamefont{and}
  \bibinfo{author}{\bibfnamefont{V.~J.} \bibnamefont{Emery}},
  \bibinfo{journal}{Nature} \textbf{\bibinfo{volume}{393}},
  \bibinfo{pages}{550} (\bibinfo{year}{1998}).

\bibitem[{\citenamefont{Emery et~al.}(2000)\citenamefont{Emery, Fradkin,
  Kivelson, and Lubensky}}]{EmeryPRL00}
\bibinfo{author}{\bibfnamefont{V.~J.} \bibnamefont{Emery}},
  \bibinfo{author}{\bibfnamefont{E.}~\bibnamefont{Fradkin}},
  \bibinfo{author}{\bibfnamefont{S.~A.} \bibnamefont{Kivelson}},
  \bibnamefont{and} \bibinfo{author}{\bibfnamefont{T.~C.}
  \bibnamefont{Lubensky}}, \bibinfo{journal}{Phys.\ Rev.\ Lett.}
  \textbf{\bibinfo{volume}{85}}, \bibinfo{pages}{2160} (\bibinfo{year}{2000}).

\bibitem[{\citenamefont{Prelov\u{s}ek et~al.}(2001)\citenamefont{Prelov\u{s}ek,
  Tohyama, and Maekawa}}]{Prelovsek01}
\bibinfo{author}{\bibfnamefont{P.}~\bibnamefont{Prelov\u{s}ek}},
  \bibinfo{author}{\bibfnamefont{T.}~\bibnamefont{Tohyama}}, \bibnamefont{and}
  \bibinfo{author}{\bibfnamefont{S.}~\bibnamefont{Maekawa}},
  \bibinfo{journal}{Phys.\ Rev.\ B} \textbf{\bibinfo{volume}{64}},
  \bibinfo{pages}{052512} (\bibinfo{year}{2001}).

\end{thebibliography}
\newpage
\else

\newpage
\fi

\ifDC
\else
\TableOne
\FigureOne
\FigureTwo
\FigureThree
\FigureFour
\FigureFive
\FigureSix
\FigureSeven
\FigureEight
\FigureNine
\FigureTen
\FigureEleven

\fi

\end{document}